\documentclass[reprint, pra, amsmath,amssymb,twocolumn]{revtex4-1}
\usepackage{graphicx,amsmath,epsfig,latexsym}

\begin{document}

\author{Ming-Guang Hu, Ruth S. Bloom, Deborah S. Jin}
\affiliation{JILA, NIST and University of Colorado, Boulder, CO 80309, USA}
\affiliation{Department of Physics, University of
Colorado, Boulder, CO 80309, USA}
\author{Jonathan M. Goldwin}
\affiliation{Midlands Ultracold Atom Research Centre, School of Physics and Astronomy, University of Birmingham, Edgbaston, Birmingham B15 2TT, UK}

\title{Search for ``avalanche mechanism" loss at an atom-molecule Efimov resonance}
\date{\today}

\begin{abstract}
The ``avalanche mechanism" has been used to relate Efimov trimer states to certain enhanced atom loss features observed in ultracold atom gas experiments. These atom loss features are argued to be a signature of resonant atom-molecule scattering that occurs when an Efimov trimer is degenerate with the atom-molecule scattering threshold. However, observation of these atom loss features has yet to be combined with the direct observation of atom-molecule resonant scattering for any particular atomic species. In addition, recent Monte-Carlo simulations
were unable to reproduce a narrow loss feature. We experimentally search for enhanced atom loss features near an established scattering resonance between $^{40}$K$^{87}$Rb Feshbach molecules and $^{87}$Rb atoms.
Our measurements of both the three-body recombination rate in a gas of $^{40}$K and $^{87}$Rb atoms and the ratio of the number loss for the two species do not show any broad loss feature and are therefore inconsistent with theoretical predictions that use the avalanche mechanism.

\end{abstract}

\maketitle

\section{Introduction}
\begin{figure}
\includegraphics[width=3in]{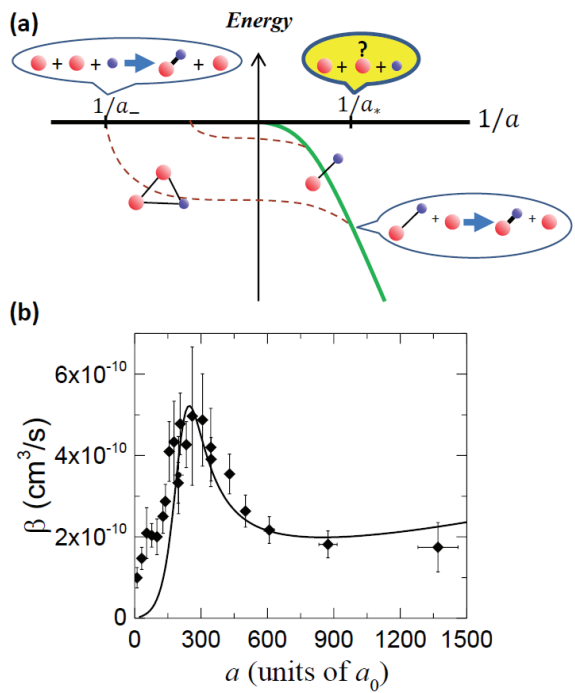}
\caption{ (Color online) Efimov loss processes. (a) Schematic showing the location of Efimov loss features. The thick black line corresponds to the threshold energy of three free atoms, the green solid line corresponds to the threshold for a KRb Feshbach molecule plus a free Rb atom, and the brown dashed lines correspond to KRb$_2$ Efimov bound states. At $a_-<0$, resonant enhancement of three-body recombination is observed. At $a_*>0$, resonant enhancement of atom-molecule inelastic collisions is observed and enhanced atom loss due to the avalanche mechanism has been postulated. In the scattering process cartoons, red and blue collision partners represent $^{87}$Rb and $^{40}$K atoms, respectively. (b) Measured atom-molecule loss rate coefficient as a function of $a$ in a mixture of Rb and RbK \cite{Bloom2013}. Resonant atom-molecule loss was observed near $a_*=230(30)a_0$; the line shows a fit to a theoretical lineshape \cite{Bloom2013,Helfrich2010}.}
\label{fig:Efimov}
\end{figure}

Evidence for Efimov three-body bound states, which were proposed originally in the context of nuclear physics \cite{Efimov1971}, has been observed in a number of ultracold atom gas experiments \cite{Kraemer2006,Ottenstein2008,Huckans2009,Zaccanti2009,Barontini2009,Pollack2009,Nakajima2010,Gross2010,Berninger2011,Wild2012,Bloom2013,Lompe2010,Knoop2009}.   In principle, near a magnetic-field Feshbach resonance \cite{Chin2010} there exists an infinite number of three-body bound states that follow a discrete scaling law. The primary signature of these three-body states in cold atom gases has been resonantly enhanced three-body loss of trapped atoms.  A loss resonance occurs at a negative value of the two-body scattering length $a$, which is denoted $a_-$, where the energy of the Efimov state coincides with the scattering threshold energy for three atoms \cite{Braaten2007}, as shown schematically in Fig. \ref{fig:Efimov} (a). Several experiments have observed multiple Efimov loss features whose locations follow discrete scaling, with each $a_-$ larger than that of the last by a factor of $e^{\pi/ s_0}$, where $s_0$ is a universal parameter \cite{Huang2014,Tung2014,Pires2014}.

An additional signature of Efimov states can be found when the energy of an Efimov state coincides with the threshold scattering energy for a Feshbach molecule and an atom.  This occurs at a positive value of $a$ denoted $a_*$ and results in resonant collisional loss in a trapped gas mixture of Feshbach molecules and atoms. Atom-molecule loss resonances have been observed for $^6$Li \cite{Lompe2010,Nakajima2010}, $^{133}$Cs \cite{Knoop2009}, and the mixture of $^{40}$K and $^{87}$Rb \cite{Bloom2013}. In addition, unanticipated resonances in the loss of trapped atoms at positive $a$ values, without initially creating molecules, have been seen for  $^7$Li \cite{Pollack2009,Machtey2012_1}, $^{39}$K \cite{Zaccanti2009}, and the mixture of $^{41}$K and $^{87}$Rb \cite{Barontini2009}. The observed loss features are relatively small, with the increase in the atom loss rate ranging from about a factor of two to a factor of five.  The features can be quite narrow, with the width ranging from a few $a_0$ to a few hundred $a_0$, where $a_0$ is the Bohr radius. These resonances are believed to be related to $a_*$ and have been attributed  to an avalanche mechanism \cite{Schuster2001,Zaccanti2009,Machtey2012_2}, whereby Feshbach molecules that are produced by non-resonant three-body recombination eject atoms from the trap via resonant, secondary atom-molecule collisions. However, there has not yet been an observation of an avalanche feature and an atom-molecule loss resonance in the same system.  In addition, a recent theoretical simulation suggests that the avalanche mechanism fails to produce a narrow atom loss feature near the atom-dimer resonance \cite{Langmack2012}.

The observation of an atom loss feature does not require the preparation of Feshbach molecules and therefore can be a simpler method for experimentally locating $a_*$.  However, it is important to verify the connection of observed avalanche peaks with atom-molecule Efimov resonances. In previous work \cite{Bloom2013}, we measured an atom-molecule loss resonance for $^{40}$K$^{87}$Rb Feshbach molecules and $^{87}$Rb atoms, but did not see any corresponding loss feature for an atom gas prepared without creating a population of trapped Feshbach molecules. The measured atom-molecule loss rate coefficient $\beta$ is presented in Fig. \ref{fig:Efimov} (b), showing resonant loss around $a_*=230(30)\,a_0$. Because this data could have missed a narrow or a small amplitude avalanche peak, we present here additional atom loss measurements on the positive $a$ side of the  $^{40}$K$-^{87}$Rb Feshbach resonance.  In particular, in order to search for an Efimov-related avalanche feature, we take many more data points with a finer spacing in $a$. In addition, we ensure uniformity of the initial atom gas conditions as we change $a$, since variation of the densities or temperature could shift or broaden a resonance feature \cite{Machtey2012_2,Langmack2012}.  Finally, we look for features in both the atom loss as well as the ratio of the number loss for Rb and K, since the avalanche mechanism should result in additional loss of Rb atoms from the resonant, secondary collisions.

The rest of paper is organized as: Sec. \ref{sec-II} describes how we prepare the ultracold Bose-Fermi mixture and measure atom loss, Sec. \ref{sec-III} presents the experimental results, which are compared against predictions based on the avalanche mechanism near an atom-molecule Efimov resonance, and Sec. \ref{sec-IV} gives conclusions.

\section{Loss measurements}\label{sec-II}
Our measurements start with an ultracold mixture of bosonic $^{87}$Rb atoms in the $|f,m_f\rangle$=$|1,1\rangle$ state and fermionic $^{40}$K atoms in the $|9/2,-9/2\rangle$ state, where $f$ corresponds to the atomic angular momentum and $m_f$ is its projection. An $s$-wave Feshbach resonance is used to control the interactions between $^{87}$Rb and $^{40}$K atoms, where $a$ as a function of magnetic field $B$ is given by $a$=$a_\text{bg}(1$-$\frac{\Delta}{B-B_\text{0}})$, $a_\text{bg}$=$-187\,a_\text{0}$, $B_\text{0}$=$546.62\,\mathrm{G}$, $\Delta$=$-3.04\,\mathrm{G}$~\cite{Klempt2008}. The atom gas is initially prepared at a magnetic field $2.07$ G below $B_0$, which corresponds to $a=88\,a_0$. We keep the temperature $T$ of the gas greater than $1.4\,T_\text{c}$ as well as greater than $0.7\,T_\text{F}$, where $T_\text{c}$ is the transition temperature for Bose-Einstein condensation of  $^{87}$Rb and $T_\text{F}$ is the Fermi temperature of $^{40}$K.

We have investigated atom loss in a single-beam optical dipole trap characterized by trapping frequencies for Rb of
$\omega_\text{r}/2\pi=600\,\mathrm{Hz}$ radially and $\omega_\text{z}/2\pi=6\,\mathrm{Hz}$ axially. In our far-detuned optical dipole trap, the trapping frequencies for K are larger than those for Rb by a factor of
1.4. The optical trap beam propagates along a horizontal direction, with a beam waist of $20\,\mathrm{\mu m}$ and a wavelength of $1090\,\mathrm{nm}$.  The atom gas mixture is prepared with an initial number of Rb atoms, $N_\text{Rb,i}$, between $7.0\times10^5$ and $9.5\times10^5$, an initial number of K atoms, $N_\text{K,i}$, between $2.6\times10^5$ and $3.9\times10^5$, and an initial temperature between 0.7 $\mu$K and 1.0 $\mu$K.

Three-body recombination produces a Feshbach molecule with a kinetic energy determined by the binding energy. In order to make sure that our trapping potential does not confine these KRb molecules or the energetic atoms resulting from scattering with KRb molecules, we want the trap depth to be lower than the binding energy of KRb molecules \cite{Zirbel2008}. As a consequence, we take measurements for $a<900\,a_0$ in a trap just deep enough to hold an atom gas with a temperature $T_{\rm{max}}=1.2\,\mathrm{\mu K}$.
 To extend our measurement to larger values  of $a$ where the binding energy of KRb is smaller, for $900\,a_0<a<1500\,a_0$ we lower our trap depth to be just deep enough to hold an atom gas with a temperature of $T_{\rm{max}}=1.0\,\mathrm{\mu K}$,
with $\omega_\text{r}/2\pi=500\,\mathrm{Hz}$ radially and $\omega_\text{z}/2\pi=5\,\mathrm{Hz}$ axially. For all of the data, the binding energy of the molecules is greater than 1.5 times $k_BT_{\rm{max}}$, where $k_B$ is the Boltzmann constant.

\begin{figure}
\includegraphics[width=3in]{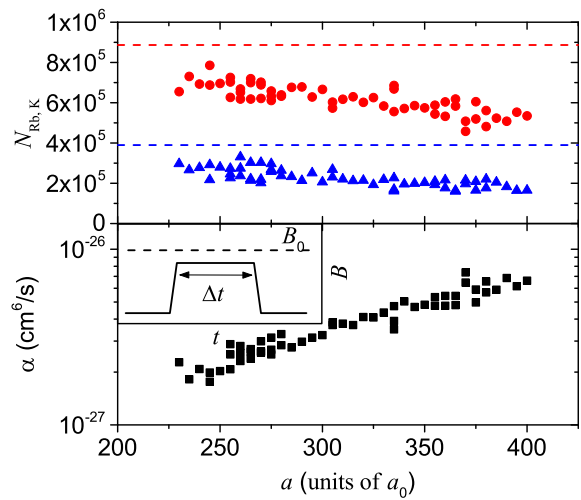}
\caption{(Color online) An example of data from which we extract the three-body recombination rate. The upper panel shows the measured atom number (circles for Rb and triangles for K) after holding at a scattering length $a$ for $1$ s. The dashed lines show the measured initial Rb and K atom numbers. The lower panel shows the extracted three-body recombination rate coefficient $\alpha$ based on Eq. (\ref{eq:alpha}). Inset shows magnetic-field sweep: the magnetic field $B$ is increased to a value near the Feshbach resonance in $0.25$ ms, held at that value for a time $\Delta t$, and then swept back to the original value in $0.25$ ms.}
\label{fig:alpha}
\end{figure}

To measure loss, we use a magnetic-field sweep to quickly increase $a$ and then wait for fixed amount of time $\Delta t$ as shown in the inset of Fig. \ref{fig:alpha}. The magnetic field is then returned to the original value where $a=88\,a_0$ and both atom species are imaged a few milliseconds after release from the optical trap.  The final atom numbers for Rb and K, which we denote $N_{\text{Rb,f}}$ and $N_{\text{K,f}}$, respectively, are determined from fits to Gaussian distributions. Combining this with the measured initial atom numbers, $N_{\text{Rb,i}}$ and $N_{\text{K,i}}$, yields the loss rate. We take data for values of $B$ during the hold time $\Delta t$ that correspond to $a$ from $100\,a_0$ to $1500\,a_0$. The hold time $\Delta t$ is changed for different ranges of $a$ in order to keep the fractional number loss $(N_\text{f}-N_\text{i})/N_\text{i}$ between $10\%$ and $60\%$, where $N_f=N_{\text{Rb,f}}+N_\text{K,f}$ and $N_i=N_{\text{Rb,i}}+N_{\text{K,i}}$. The value of $\Delta t$ varies from $5\,\mathrm{s}$ at small $a$ to $2.5\,\mathrm{ms}$ at large $a$.

Fig. \ref{fig:alpha} shows a subset of our loss measurement data for a hold time, $\Delta t$, of 1 s. This two-point measurement approach (measuring the number at time 0 and at time $\Delta t$) trades accuracy for precision.  Specifically, a full measurement of the loss curve, where the atom numbers are measured at many different times, allows for a more accurate determination of the three-body rate coefficient at a particular value of $B$, however the faster two-point measurement minimizes the effect of drifts in experiment parameters and therefore enhances the precision and our ability to detect any small loss peaks as we vary $B$.

In order to combine data taken for different hold times $\Delta t$, we extract an approximate three-body rate coefficient $\alpha$. For three-body recombination of $^{87}$Rb+$^{87}$Rb+$^{40}$K, $\alpha$ is defined by
$\dot{N}(t)=-3\alpha \int d^3\mathbf{r} n_\text{K}(\mathbf{r},t)n_\text{Rb}^2(\mathbf{r},t)$ \cite{Bloom2013}, where $n_\text{Rb}(\mathbf{r},t)$ and $n_\text{K}(\mathbf{r},t)$ are number densities of $^{87}$Rb and $^{40}$K, respectively. To simplify this differential equation, we can use the fact that K and Rb share almost the same polarizability in our optical dipole trap and ignore the small relative sag between Rb and K clouds.
Assuming a Gaussian density profile consistent with a harmonically trapped Maxwell-Boltzmann gas, we can re-write the integral in terms of the total number $N$ and temperature $T$ as
$\dot{N}(t)=-3\alpha A\bar{\omega}^6N^3/T^3$ \cite{Weber2003}.
Here, $\bar{\omega}=({\omega_\text{r}}^2\omega_\text{z})^{1/3}$, $A=\frac{R^2}{(1+R)^3}\left(\frac{m_{\text{Rb}}}{2 \pi \sqrt{3} k_B} \right)^3$, $R$ is the number ratio $N_{\text{Rb}}/N_{\text{K}}$, $m_{\text{Rb}}$ is the atom masses of $^{87}$Rb. Although the number ratio $R$ can change during a measurement, the parameter $A$ is only weakly dependent on $R$. In the approximation that the temperature and the parameter $A$ are constant during $\Delta t$, $\alpha$ can be solved for analytically,
\begin{equation}\label{eq:alpha}
\alpha=\left[\frac{1}{N_\text{f}^2}-\frac{1}{N_\text{i}^2}\right]\frac{T^3}{6A\bar{\omega}^6\Delta t}.
\end{equation}
Using the average initial number ratio $R=N_{\text{Rb,i}}/N_{\text{K,i}}=2.5$ and initial temperature $T$, we obtain $\alpha$ using Eq. (\ref{eq:alpha}) (see Fig. \ref{fig:alpha}, lower panel).  As a check, we have compared the results from our two-point measurements using Eq. (\ref{eq:alpha}) with previous data where $\alpha$ was extracted from many measurements of the number of atoms as function of time \cite{Bloom2013}, and we find that they agree to within a factor of $2$.

\section{Experimental results}\label{sec-III}
We previously identified an Efimov-like resonance between $^{40}$K$^{87}$Rb Feshbach molecules and $^{87}$Rb atoms at the scattering length $a_*=230\,a_0$ with width of roughly $200\,a_0$ \cite {Bloom2013}. According to Refs. \cite{Zaccanti2009,Machtey2012_2}, this atom-molecule resonance can also result in enhanced atom loss at scattering lengths near $a_*$ for a gas initially consisting of atoms only.
Here, non-resonant three-body recombination of atoms produces energetic $^{40}$K$^{87}$Rb Feshbach molecules that then collide with atoms multiple times to result in atom loss.
In addition to enhanced total atom loss, our two-species atom gas could provide an additional signature for this avalanche scenario. Namely, the number loss ratio $\Delta N_\text{Rb}/\Delta N_\text{K}$ should also show a resonant increase that coincides with the enhanced atom loss feature, since the collision channel $^{40}$K$^{87}$Rb+$^{87}$Rb is enhanced while $^{40}$K$^{87}$Rb+$^{40}$K is not \cite {Bloom2013}.  Here, the number loss ratio is defined by
\begin{equation}\label{eq:ratio}
  \Delta N_{\text{Rb}}/\Delta N_{\text{K}}=\frac{N_\text{Rb,f}-N_\text{Rb,i}}{N_\text{K,f}-N_\text{K,i}}.
\end{equation}

\begin{figure*}
\includegraphics[width=5in]{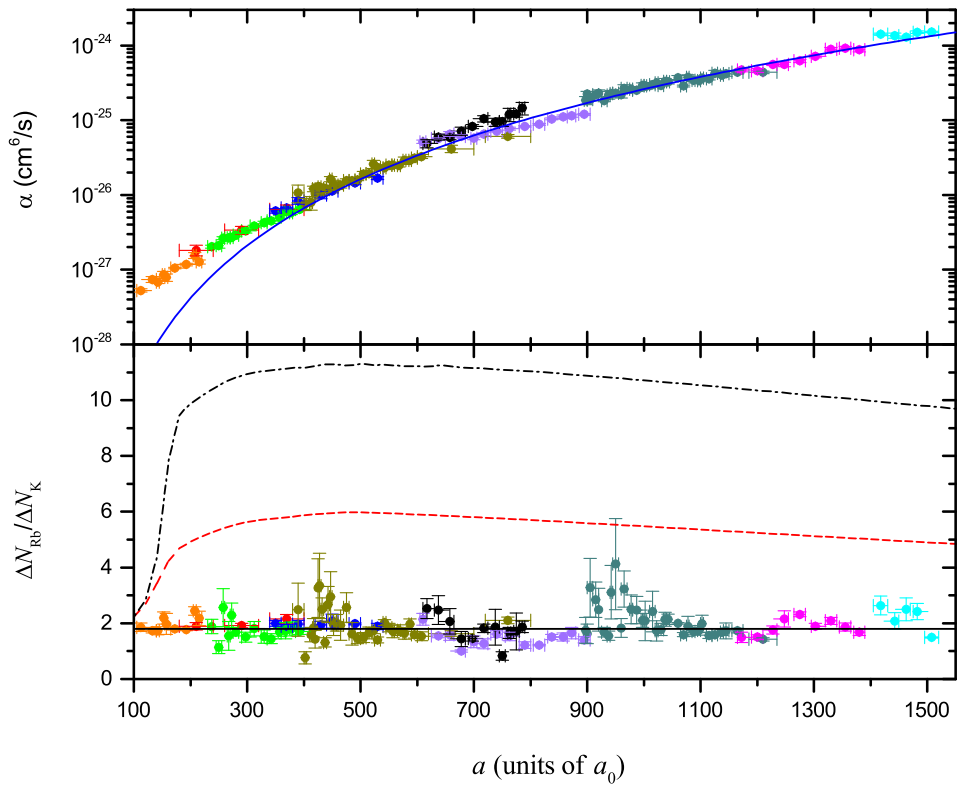}
\caption{(Color online) Atom loss measurements in a single-beam optical trap. Different datasets are color-coded and have different values of the holding time $\Delta t$ ranging from 5 s to 2.5 ms. The upper panel shows the measured three-body recombination rate coefficient $\alpha$ (points) versus $a$. The blue solid line indicates an $a^4$ dependence, which is expected in the absence of Efimov resonances. The lower panel shows the measured number loss ratio $\Delta N_{\text{Rb}}/\Delta N_{\text{K}}$ (points) versus $a$. The black solid line corresponds to the average value of $1.8$. The red dashed line and black dot-dashed line come from calculations based on a probability model from Ref. \cite{Machtey2012_2} with $\eta_*=0.26$ and $\eta_*=0.02$, respectively.   }
\label{fig:avalanche1}
\end{figure*}

Fig. \ref{fig:avalanche1} shows our measurement results, where each point on the plots shows the average of four repeated measurements within each dataset having the same value of $\Delta t$. The vertical error bars indicate the standard deviation of the mean, while the horizontal error bars indicate the range of $a$ values used in the averaging.  The upper panel in Fig. \ref{fig:avalanche1} shows the loss rate coefficient $\alpha$ extracted using Eq.~(\ref{eq:alpha}) while the lower panel shows the number loss ratio $\Delta N_\text{Rb}/\Delta N_\text{K}$.  We see no clear evidence for an avalanche peak.  Specifically, aside from the deviation from $a^4$ scaling at small values of $a$, the dominant features in $\alpha$ appear to be small systematic shifts that occur when we combine datasets taken with different values of $\Delta t$, and we can easily rule out the presence of any feature where $\alpha$ is increased by a factor of two or more. The measured number loss ratio $\Delta N_\text{Rb}/\Delta N_\text{K}$ has an average value of approximately $2$, which is the expected value for three-body recombination with no additional avalanche mechanism loss. The measurement of $\Delta N_\text{Rb}/\Delta N_\text{K}$ has a lower signal-to-noise ratio than $\alpha$ and one can identify some possible peaks in the data.  However, these peaks have no corresponding feature in $\alpha$.  In addition, our measured number loss ratio is qualitatively inconsistent with predictions from an avalanche mechanism model as shown by dashed and dot-dashed lines in the lower panel of Fig. \ref{fig:avalanche1}. The amplitude of these potential peaks in our data is smaller than that of the model by a factor of 2 or more and the width is narrower by a factor of 10 or more. The avalanche model is described in detail below.

In Ref. \cite{Machtey2012_2}, Machtey \emph{et al.} present an improved avalanche model based on Ref. \cite{Zaccanti2009}. The results of this model show qualitative agreement with the two $^7$Li experiments \cite{Pollack2009,Machtey2012_1} and the $^{39}$K experiment \cite{Zaccanti2009}, although the predicted widths for enhanced loss were typically several times larger than the observed avalanche peak widths.  Monte-Carlo methods have also been used to simulate the avalanche loss for homonuclear systems~\cite{Langmack2012}. These simulations were applied to the $^7$Li system and result in an even wider avalanche loss feature, with widths that are 10 to 20 times larger than the observed atom loss features~\cite{Langmack2012}.

For comparison with our data, we have applied the model of Machtey \emph{et al.}, which we modify for the heteronuclear case, to calculate the expected avalanche peak for our experiment parameters.  In the Machtey \emph{et al.} model, the elastic and inelastic atom-dimer cross sections are used to calculate the probability that a dimer created by three-body recombination undergoes a specific number of secondary elastic collisions with atoms before exiting the trap.  A weighted sum of these probabilities then yields the expected number of extra atoms lost due to the avalanche mechanism.  To extend this model to a two-species gas, we use the scattering length for a Rb atom and a KRb Feshbach molecule given in Ref. \cite{Helfrich2010}:
$a_\text{AM}(a)=\left[C_1+C_2\cot(s_0 \ln(a/a_*)+i\eta_*)\right]a$ with constants $C_1=1.14$, $C_2=2.08$, and $s_0=0.6536$. The Efimov parameters, $\eta_*=0.26$ and $a_*=230\,a_0$, are taken from the fit to atom-dimer trap loss data in Ref. \cite{Bloom2013}. In terms of $a_\text{AM}$, the atom-molecule elastic and inelastic cross sections are given by $\sigma_\text{el}(a)=4\pi |a_\text{AM}(a)|^2$ and $\sigma_\text{inel}(a)=-4\pi \mathrm{Im}(a_\text{AM}(a)/k)$, respectively \cite{Braaten2007}. For the initial collision, the relative wave number $k$ is calculated assuming the atom is essentially at rest and the molecule has a kinetic energy given by $\left[m_{\text{Rb}}/(m_{\text{K}}+2m_{\text{Rb}})\right]E_\text{b}$, where $E_\text{b}$ is the binding energy of the molecule. In each subsequent collision with an atom at rest, the mean energy of the dimer is multiplied by a factor of $\left[\left(m_{\text{K}}+m_{\text{Rb}}\right)^2+m_{\text{Rb}}^2\right]/\left(m_{\text{K}}+2m_{\text{Rb}}\right)^2\approx0.52$.

For secondary collisions, a key parameter is the column density of the trapped atom gas, $nl$. In the calculation, we use the average density of the Rb atoms for $n$ and the geometric mean root-mean-squared width of the trapped gas for $l$.  For our data, $n=1.1\times10^{13}\,\text{cm}^{-3}$ and $l=12\,\mu\text{m}$. With these parameters, we calculate the mean number of Rb atoms lost per three-body recombination event, which can be directly compared to our number loss ratio data. The red dashed curve in the lower panel of Fig.~\ref{fig:avalanche1} shows the model result.
Because a much lower Efimov resonance inelasticity parameter, $\eta_*$, can be extracted from fits to three-body loss data \cite{Bloom2013}, we also show the model result for $\eta_*=0.02$ (black dot-dashed curve in Fig.~\ref{fig:avalanche1}).
For either value of $\eta_*$, the model predicts a wide resonance feature in atom loss and in $\Delta N_\text{Rb}/\Delta N_\text{K}$, which is clearly inconsistent with our measurements.

Given that we do not observe a feature consistent with these predictions for an Efimov avalanche peak, it is useful to compare the parameters for our system to those of experiments where avalanche peaks have been observed.   In particular, compared to the $^7$Li experiment of Ref.~\cite{Machtey2012_1}, our temperature is very similar (within a factor of 2), the mean size of the trapped gas $l$ is similar (within $30\%$), and our atom density is an order of magnitude larger, which should be favorable for observing an Efimov avalanche peak.  In addition, the Efimov resonance parameters $a_*$ and $\eta_*$ used to model the $^7$Li  loss feature \cite{Machtey2012_2} are very similar to those for the $^{40}$K-$^{87}$Rb case.  Our trap aspect ratio is larger than that of Ref.~\cite{Machtey2012_1} by a factor of 15, but similar to that of Ref.~\cite{Pollack2009}, which also reported an avalanche peak for $^7$Li. Finally, we note that we have also taken measurements in a crossed-beam optical dipole trap with aspect ratio of 30, and again no clear avalanche loss feature was observed.

\section{Conclusions}\label{sec-IV}
We have measured the three-body recombination loss rate and the number loss ratio for a $^{40}$K-$^{87}$Rb atom gas mixture at positive scattering length over the range from $100\,a_0$ to $1500\,a_0$ in a search for a feature connected to the previously observed atom-dimer Efimov resonance at $a_*=230\,a_0$.
While an avalanche model has been used to interpret atom loss features seen in other systems as being a consequence of an atom-dimer resonance, our measurements do not show a loss feature consistent with this model. The fact that there remains no single system in which both resonant loss in an atom-dimer gas mixture and a corresponding loss feature for an atom gas have been observed is problematic for validation of this explanation of these atom loss peaks.

\begin{acknowledgements}
 This work is supported by NIST and by NSF under Grant No. 1125844. JG gratefully acknowledges support from the JILA Visiting Fellows program.
\end{acknowledgements}

\end{document}